\documentclass[12pt,preprint]{aastex}

\lefthead{Escala}
\righthead{The Similitude Principe in Biology}

\usepackage{lineno}

\begin{document}


\title{The Principle of Similitude in Biology: From Allometry to the Formulation of Dimensionally Homogenous `Laws'.}
\author{Andr\'es Escala}
\affil{Departamento de Astronom\'{\i}a, Universidad de Chile, Casilla 36-D, Santiago, Chile.}
\affil{aescala@das.uchile.cl}

\begin{abstract}
Meaningful laws of nature must be independent of the units employed to measure the variables.
 The principle of similitude (Rayleigh  1915) or dimensional homogeneity, states that only commensurable quantities (ones having the same dimension) may be compared
, therefore, meaningful laws of nature must be homogeneous equations in their various units of measurement, a result which was formalized in the $\rm \Pi$ theorem (Vaschy 1892; Buckingham 1914). However, most relations in allometry do not satisfy this basic requirement, including the 
`3/4 Law' (Kleiber 1932) that relates the basal metabolic rate and body mass, 
besides  it is sometimes claimed to be the most fundamental biological rate (Brown et al. 2004)  and the closest to a law in life sciences (West \& Brown 2004). Using the $\rm \Pi$ theorem, here we show that it is possible to construct an unique homogeneous equation for the metabolic rates, in agreement with  data in the literature. We find that the variations in the dependence of the metabolic rates on body mass are secondary, coming from variations in the allometric dependence of the heart frequencies. This includes not only different classes of animals (mammals, birds, invertebrates) but also differences during rest and exercise (basal and maximal metabolic rates). Our results demonstrate that most of the differences found in the allometric exponents (White et al. 2007) are due to compare incommensurable quantities and that our dimensionally homogenous formula, unify these differences into a single formulation. We discuss the ecological implications of this new  formulation in the context  of the Malthusian's, Fenchel's and the total energy consumed in a lifespan relations.

\end{abstract}


\section{Introduction}

The invariance of nature under scaling of units of measurement has been a powerful tool for the  discovery of new phenomena, used for centuries  in physics (Fourier 1822) and related areas (Macagno 1971). Being called   dimensional analysis or  principle of similitude (Rayleigh 1915),  this methodology have  been  also applied   to the solution of complex problems, ranging from the atmosphere behavior under  a nuclear explosion (Taylor 1950) to airfoil prototypes (Bolster et al. 2011) and the formation of stars in galaxies (Escala 2015; Utreras et al. 2016)

Scaling invariance is desirable in any branch of science, not only restricted to physics, however, in many areas the fundamental relations have not yet been formulated in a form independent of the scaling of units. 
In the  case of life sciences, particularly relevant  are the allometric scaling laws, which relates a physiological variable with body size, which in almost all cases is measured by its mass. To fulfill   the similitude  principle implies that these relations can always be rearranged in terms of non-trivial dimensionless parameters/groups
, however, allometric scalings almost never  fulfill this property, which is also called dimensional homogeneity.  Another way of formulating this issue, is as a phenomena that can be  described with homogenous equations only with the aid of as many dimensional constants  as there are variables (Bridgman 1922).

This problem is illustrated  in  the influential book `Scaling' by Schmidt-Nielsen  (1984), 
 that  shows dozens of relations between physiological variables in species   and  body mass, being the most notable one the so-called Kleiber's Law, between basal metabolic rate (energy consumption per unit time) and body mass. None of the nontrivial relations satisfies the similitude principle, including the  famous Kleiber's Law,  with one notable exception: the  allometric relation between swimming speed of fishes and tail-beat frequency (Bainbridge 1958). When the fishes swimming speed is properly normalized dividing it by their body length,  thus having frequency units as the tail-beats, the data among different fishes lies into a single curve in the same fashion as, for example, von K\'arm\'an (1957) unified different experiment of turbulent flows in pipes using only dimensional analysis.
 
The self-similarity displayed in Brainbridge's allometric relation is appealing since it is the same displayed in physics and will be the main topic of this work. However, it is important to note first that there is another partial approach to guarantee similitude: fractal geometry. The possibility of having fractional dimensions allows to satisfy similitude with the use of an extra parameter, the Haussdorf or fractal dimension D. If the fractal dimension is properly chosen, is still possible  to fulfill similitude 
for power laws with any fractional exponent.

In terms of dimensional analysis, fractal curves can be identified as incomplete similarities and  such approach is   the same used in renormalization group analysis  in  modern theoretical physics, since it can  be proven the equivalence of  incomplete similarity and invariance with respect to the renormalization group (Barenblatt 2003). 
Fractals    belongs to the group of self-similar solutions of the second kind,  in which the requirements of homogeneity and self-similarity are necessary to be satisfied only  locally (Barenblatt 2003). 
On the other hand, 
 self-similar solutions of the first kind (or complete similarities) that satisfies the more restrictive requirements of global homogeneity and self-similarity, can sometimes be obtained using the tools of dimensional analysis.

Fractals models of the allometric relations  became   popular in recent years, most notably since West et al (1997), that conjectured a universal `1/4 power law' mass scaling for  physiological variables and used fractal geometry to explain it from first principles. This gave considerable attention to fractal models for  the allometric relations, however, also debate  on the validity of such `1/4 power law' universality (see  for example  the  critical review by Hulbert 2014 or   the comparison between  theoretical predictions by  Price et al. 2009). In fact, analysis of 127 interspecific allometric exponents discarded an universal  metabolic relation (White et al. 2007) and using statistical analysis, Dodds et al. (2001)  again  discarded  a simple scaling law for metabolic rates. Due to this intense debate, we will start analyzing possible fractals models
, to then move towards  
complete similarity solutions using dimensional analysis tools ($\Pi$ theorem).

This paper 
is organized as follows. We start studying the mathematical requirements to be a natural law in \S 2. We continue using dimensional analysis to study  possible fractal models  for the metabolic rates  in \S 3. 
 Section 4 continues with the search for a dimensionally homogenous metabolic rate equation and compare it   against  data taken from the literature. 
Finally in \S 5, we discuss  the  results and implications of this work.

\section{
Dimensional Constants and Homogeneous Equations} 

In the  proper mathematical formulation of natural laws,  a minimum requirement is to be expressed in a general form that remain true when the size of units is changed, therefore, meaningful laws of nature  must be homogeneous equations in their various units of measurement, as was first observed by Fourier 
(1822). This requires, in addition to be a dimensionally homogeneous equation,  that the constants with dimensions are restricted to only universal ones and to a minimum number, which in no case could exceed  the total number of fundamental units of the problem. 

However, in many areas (allometry being  one of those) the phenomena can be described in homogenous equations only with the aid of as many dimensional constants as there are variables (Bridgman 1922). One example of such relations is the `3/4 Law' (Kleiber 1932) that relates the basal metabolic rate (B) and body mass (M), described with  an equation of the form 
\begin{equation}
\rm B = B_0 \, \Big(\frac{M}{M_0} \Big)^{3/4} \, ,
\label{Kleiber}
\end{equation} 
where $\rm B_0$ and $\rm M_0$ are normalization constants 
and therefore, in this relation each variable  has its own constant with units. The ubiquity of this type of formulation  for expressing relations,  probably  comes  from the fact that the procedure  of fitting a curve to a set of points gives such constants with dimensions,  then it is natural to express the relation found in such terms. 

The limitations of this approach  can be  better understood  analyzing  an specific problem, like the following thought experiment: let's suppose   that  we are able to perform accurate experiments about the gravitational acceleration suffered by an object of mass m at a distance h from the surface of a planet. If we perform these experiments in the Earth and if any kind of friction or dissipation is negligible, taking the same approach used in allometric relations we will find that the acceleration data can be fitted by a relation of the form $\rm a = a_0 (1 + h/h_0)^{-2}$.

If we then travel to take measurements to a different planet, Mars for example, to perform  the same experiment we will be able to fit the same `acceleration law',  $\rm a = \tilde a_0 (1 + h/ \tilde h_0)^{-2}$, but with different normalization constants $\rm \tilde{a}_0$ and $\rm \tilde{h}_0$. Same case  will happen if we take data from Jupiter, Neptune, etc. This methodology is exactly the one  of as many dimensional constants $\rm(a_0 \, \&  \, h_0$'s) as variables (a \& h's) and is the same scenario found in allometric relations (Schmidt-Nielsen 1984). Moreover, in this `one constant per variable' approach to study the `acceleration law', we have also one law per planet studied, as in allometric relations where we have one law for each different classes of animals, with well documented intra-specific variations for invertebrates versus vertebrates, endotherms vs ectotherms, running vs resting, etc (White et al. 2007). 

It is important to compare now the just mentioned methodology, to the one that lead to the most general `acceleration law' of this problem. Thanks to Newton we  know the  general answer, not only for the Earth, but for any given spherical planet of mass M and radius R, which is given by
\begin{equation}
\rm a = \frac{||\bold{F_g}||}{m} = \frac{GM}{R^2} \times \Big(  1+ \frac{h}{R} \Big)^{-2} \, ,
\label{Newton}
\end{equation} 
where G is the universal constant of gravitation. This solution comes from two well formulated laws,  expressed by homogeneous equations  in a form independent of the units of measurements: Newton's Law of Universal Gravitation and his 2nd Law of motion. Such laws, were conjectured by Newton motivated on pure empirical observations  and from them, a theory was constructed that explained a whole variety of phenomena, in which the problem described by Eq. \ref{Newton}, is just one of many. In terms of dimensional analysis, this approach to the gravitational free-fall problem has 3 relevant dimensions (mass, length and time) and only one universal constant with units (G), compared to the previous case, in which as many constants as twice the number planets  are needed to describe the same phenomena. 

From the comparison of both approaches, the single well formulated dimensionally homogeneous solution (Eq. \ref{Newton}) and  the one with as many normalization constants ($\rm a_0$ and $\rm h_0$'s) as variables (a and h's), it is clear that this multiplicity of  laws and normalizations  with units, are needed to explain variations of other relevant variables  of the problem (R and M in this case). Also, it is straightforward to wonder if many of the exceptions in the Kleiber's law and other allometric relations, are hiding variations of other relevant variables  in the metabolic rate relation.

The difference between a dimensionally homogenous law that fulfill the similitude principle and phenomena described with the aid of as many dimensional constants as there are variables, is well known and used for centuries in physics. Newton  was probably the first that realized this difference, when he conjectured his laws of motion and gravitation to explain pure empirical relations like the  ones studied by Kepler, to describe the planetary motions around the Sun. Fourier   was the first to realize the importance of units and that the laws must be homogeneous equations in their various forms. Maxwell established  modern use of dimensional analysis in physics by distinguishing mass, length, and time as fundamental  units, while referring to other units as derived. Contemporary physics allows a maximum  of 3 fundamental  constants  (speed of light, Planck's constant and G; Duff et al. 2002), which combined defines the natural base system to express the 3 fundamental units,  called the Planck's mass, length and time.

 The rules required in the mathematical formulation of any natural law, was eventually formalized in the $\Pi$ theorem discovered by Vaschy (1892), Buckingham (1914) and others. We will be use such theorem in the next sections, to search for a dimensionally homogenous law for the metabolic rates, starting with the less restrictive incomplete similarity solutions (fractal models) to then  continue studying  complete similarity ones.

\section{Dimensional Analysis  of Fractal Models}

The Vaschy-Buckingham $\rm\Pi$ theorem defines the rules to be fulfilled by any meaningful  relation aimed to be law of nature and it is a formalization of Rayleigh's similitude principle. The theorem states that if there is a  meaningful equation involving a certain number, n, of  variables, and k is the number of 
relevant dimensions, then the original expression is equivalent to an equation involving a set of p = n $-$ k  dimensionless parameters constructed from the 
original variables. Mathematically speaking, if we have the following  equation:
\begin{equation}
F(A_1,A_2,\ldots,A_n)=0 \, ,
\label{V-B}
\end{equation} 
where the Ai are the n  variables that  are expressed in terms of k independent units, Eq \ref{V-B} can be written as
\begin{equation}
f(\Pi_1, \Pi_2, \ldots, \Pi_{n-k})=0\,  ,
\label{V-B2}
\end{equation} 
where the $\rm \Pi_{i}$ are dimensionless parameters constructed from the $\rm A_i$ by p = n $-$ k dimensionless equations  of the form $\rm \Pi_i=A_1^{m_1}\,A_2^{m_2}\cdots A_n^{m_n}$. 
 
We will 
 follow the more general approach of Barenblatt \& Monin (1983),  to study a possible the fractal-like nature of biological systems, since it do not depend on a particular mechanistic model  like the one in West et al. (1997), or   assumes  that the  basal metabolic rate ($\rm \dot{V}_{O_2} $) is directly proportional to the effective fractal surface of the body, as in West et al. (1999). We can construct a model by  assuming  that the metabolic rate, which is commonly  measured in milliliters of $\rm O_2$ per minute and has dimensions $\rm [\dot{V}_{O_2}]=[M_{O_2}]/[T]$, only depends on an absorbing capacity $\rm \beta_D$ with dimensions $\rm [ \dot{V}_{O_2}][L]^{-D}$, the body mass W with dimensions [M] and the density $\rm \rho$ with dimensions $ [M][L]^{-3}$. 
 The absorbing capacity $\rm \beta_D$  characterizes if oxygen-absorbing part of an organ could be approximated by a line (D=1), by  a surface (D=2),  by a volume (D=3)) or by an intermediate fractional dimension D (for more details, see Barenblatt \& Monin 1983)

Because this model has n=4  variables ($\rm \dot{V}_{O_2} $, $\rm \beta_D$, W \&  $\rm \rho$) and k=3 independent units ($\rm [\dot{V}_{O_2}]$, [M], [L]),  the $\Pi$ theorem  tells that one dimensionless parameter can be constructed, since n-k=1. To find the
dimensionless parameter  is straightforward by looking integer exponents such $\rm \Pi_{1} =  \dot{V}_{O_2} \beta_D^{a} W^{b}  \rho^{c}$ has no dimensions. This is equivalent to force $\rm [\dot{V}_{O_2}]^{a+1}$ $\rm [M]^{b+c}$ $\rm [L]^{-aD-3c}$ to  be dimensionless,  which gives a set of 3 algebraic equations (a+1=0; b+c=0; aD+3c=0) that has the  following solution: $\rm a =-1$, $\rm b=-D/3$ and $\rm c=D/3$. Therefore, for this model it  is possible to construct the following dimensionless quantity (see Barenblatt \& Monin 1983 for details):

\begin{equation}
\rm \Pi_1 =  \frac{\dot{V}_{O_2} }{\rm \beta_D (W/\rho)^{D/3}}  \, ,
\label{pi}
\end{equation} 
which implies an  allometric scaling for the  oxygen consumption rate of $\rm \dot{V}_{O_2}  = A W^{D/3}$, with A = $\rm \Pi_1 \beta_D \rho^{-D/3}$.  For a fractional dimension D=2.25, the usual 3/4 exponent of the Kleiber's Law is recovered (Barenblatt \& Monin 1983; Turcotte et  al. 1998). 

Another interesting quantity to look is the specific metabolic rate, $\rm \dot{V}_{O_2} $/W, which from Eq.  \ref{pi} takes the form:
\begin{equation}
\rm  \frac{\dot{V}_{O_2}}{W} = \frac{\Pi_1   \beta_D W^{\frac{D}{3}-1}}{\rho^{D/3}} \propto W^{\frac{D}{3}-1}\, .
\label{specific}
\end{equation}

Since $\rm  \dot{V}_{O_2}/W$ has  dimensions of $\rm [M_{O_2}][T]^{-1}[M]^{-1}$, it can be rearranged as equals to $\rm  \Pi_1  \eta_{O_2}  \nu_o $, where   $\rm \eta_{O_2}$ is an specific $\rm O_2$ absorption factor, with units  $\rm [M_{O_2}]/[M]$ and  $\rm \nu_o$ that can  be identified as a characteristic frequency, since it has inverse time units ([T]$^{-1}$).  Assuming  an absorption factor $\rm \eta_{O_2}$  independent of W, the characteristic frequency scales with body mass as $\rm \nu_o \propto W^{\frac{D}{3}-1}$, which can be directly compared against  relevant rates  like the respiratory and/or heart frequencies, in which there is considerable data on its scaling with W.  This gives an independent Hausdorff  dimension   value needed to fulfill  dimensional homogeneity. 

For the case of mammals, heart frequencies scales as $\rm W^{-0.25}$ (Brody 1945, Stahl 1967) which for a $\rm \nu_o \propto W^{\frac{D}{3}-1}$, implies a fractional dimension of D=2.25, in agreement with the fractional dimension determined from the Kleiber's Law. A similar result is obtained if we instead use, as characteristic frequency, the respiratory rates of mammals at rest (Calder 1968). In the case of sub groups, for example marsupials mammals have metabolic rates $\rm \propto M^{0.74}$  and heart frequencies $\rm \propto M^{-0.27}$, that  gives in both cases a dimension of D$\sim$2.2 from Eqs. \ref{pi} \& \ref{specific} respectively. In invertebrates such as spiders, the metabolic rates  scales as $\rm M^{0.59}$ (Anderson 1970, 1974) and heart frequencies scales as $\rm M^{-0.41}$ (Carrel \& Heathcote 1976), quite different that in mammals, but  both the metabolic rates  and heart frequencies  implies the same fractional dimension of D$\sim$1.8. In birds the scaling is again similar to mammals, proportional to $\rm M^{0.72}$ (Lasiewski \& Dawson (1967) and $\rm M^{-0.23}$ (Calder 1968), giving slightly different dimensions of D = 2.2 and 2.3 respectively.

We see a general consistency between the fractional dimensions D determined independently from the metabolic rates and  heart/respiratory  frequencies, regardless of the considerable variations of the D value among groups  (especially in the case of invertebrates). It can be argued that such differences, are natural due to the different  evolutionary stages  among animal groups. Nevertheless, in the case of invertebrates is in addition required  to    argue in favor of oxygen-absorbing organs that are better approximated by fractal lines (D$<$2) than  by fractal surfaces (2$\leq$D$\leq$3).

Another of the major issues in the  exponent of the metabolic rate, is the change of the allometric scaling in the oxygen consumption under exercise conditions (Weibel 2002), or maximal metabolic rate $\rm \dot{V}_{O_2}^{max}$. The scaling for $\rm \dot{V}_{O_2}^{max}$ is  approximately  proportional to $\rm M^{0.85}$ (with slope variations ranging from 0.83 to 0.88; Savage et. al 2004,  Taylor et al 1981, Dlugosz et al 2013, Weibel and Hoppeler 2005 and  Bishop 1999) and the heart frequencies under such $\rm \dot{V}_{O_2}^{max}$ conditions, scales as $\rm M^{-0.15}$ (Weibel and Hoppeler 2004, 2005). Again, both scalings gives the same fractional dimension, which in this case is D= 2.55. However, this kind of models based on fractal geometry requires now a change from D=2.25 to 2.55, in order to explain the change of allometric scaling under exercise conditions. This assumption of time-dependent changes of the fractal network have not yet been observed, although it opens an interesting new possibility. 


The requirement  of adjustable fractal network under different exercise conditions is independent of the particular fractal model studied, unless the model 
assumes that the metabolic rate  is directly proportional to the effective fractal surface of the animal, like in  West et. al (1999), in which  is not  possible to explain the  $\rm \propto M^{0.85}$ scaling of the $\rm \dot{V}_{O_2}^{max}$. In such a case, the maximal possible scaling is $\rm \propto M^{0.75}$, which correspond to  volume filling surfaces (West et al. 1999).


In  the different cases studied, we do not found a single and universal Hausdorff dimension D that explains  allometric scaling for all organisms, but we found an excellent agreement in the  dimension D determined from two different empirical allometric relations (metabolic rate and frequency). Nevertheless, the fractal model studied 
 also assumes   that the characteristic frequency $\rm \nu_o$ (Eq. \ref{specific}) has all the D dependence on the W exponent, therefore in this particular model  the variations in the W  scaling of  the $\rm \dot{V}_{O_2} $ are secondary, coming directly  from its dependence on  $\rm \nu_o$. The agreement between the model and allometric relations, therefore suggests that indeed  the W  dependence in $\rm \dot{V}_{O_2} $ and $\rm \nu_o$ might not be independent. 

If  such scenario ends up being the case for  the metabolic rate relation, we are dealing with an analogous  case to the acceleration example studied in \S 2, where  multiple   laws and normalizations were needed to explain variations of other relevant variables of the free-fall problem (R and M). In the metabolic rate relation, the  multiplicity of W scalings could be in fact needed to explain variations of another relevant variable: the characteristic frequency $\rm \nu_o$
. This  motivates us to explore, in the next section,  $\rm \nu_o$ as an independent physiological variable that controls the metabolic rate  $\rm \dot{V}_{O_2} $.

\section{A Complete Similarity Solution}

Assuming now that   the characteristic frequency $\rm \nu_o$  is an independent  variable in  the metabolic rate relation, in this section we  have  a different model with n=4 independent  variables ($\rm \dot{V}_{O_2} $, $\rm \eta_{O_2}$, W \&  $\rm \nu_o$) and  that  has now  k=3 independent units ($\rm [M_{O_2}]$, [M], [T]), therefore, from the $\Pi$ theorem  we know  that n-k=1 dimensionless parameters can be constructed. The  dimensionless parameter  is again found by looking integer exponents such $\rm \Pi_{1} = \dot{V}_{O_2}  \eta_{O_2}^{a} W^{b} \nu_o^{c}$ has no dimensions,  which has the  unique solution: $\rm a = b = c = -1$. This implies that the dimensionless parameter is $\rm \Pi_{1}  = \dot{V}_{O_2}  \eta_{O_2}^{-1}W^{-1} \nu_o^{-1} $ and if the metabolic rate  
relation depends only on this four  variables, the Vaschy-Buckingham $\rm \Pi$ theorem states that it should be  a function f such $\rm f(\Pi_{1}  = \dot{V}_{O_2} \eta_{O_2}^{-1} W^{-1} \nu_o^{-1} )=0$ (Eq. \ref{V-B2}). If function f have a zero that we called $\rm \epsilon$, such f($\rm \epsilon$)=0,  implies:
\begin{equation}
\rm  \dot{V}_{O_2}  =  \epsilon \eta_{O_2}  \nu_o W \, ,
\label{new}
\end{equation}
which is a self-similar solution of the first kind or complete similarity.

In order to confirm that the variations of the  W  scaling in the metabolic rate   $\rm \dot{V}_{O_2}$ comes  from its dependence on $\rm \nu_o$, 
the self-similar solution found (Eq. \ref{new}) needs to be contrasted  against empirical data. For that, it is required  to  check if  the  dimensionally homogenous relation  has the   slope  of an identity, which must be unity within errors. For that purpose, we collected metabolic rates, masses and characteristic  frequencies  for different groups  (mammals and birds) and exercise conditions (basal and maximal). 

\begin{figure}[h!]
\begin{center}
\includegraphics[width=11.9cm]{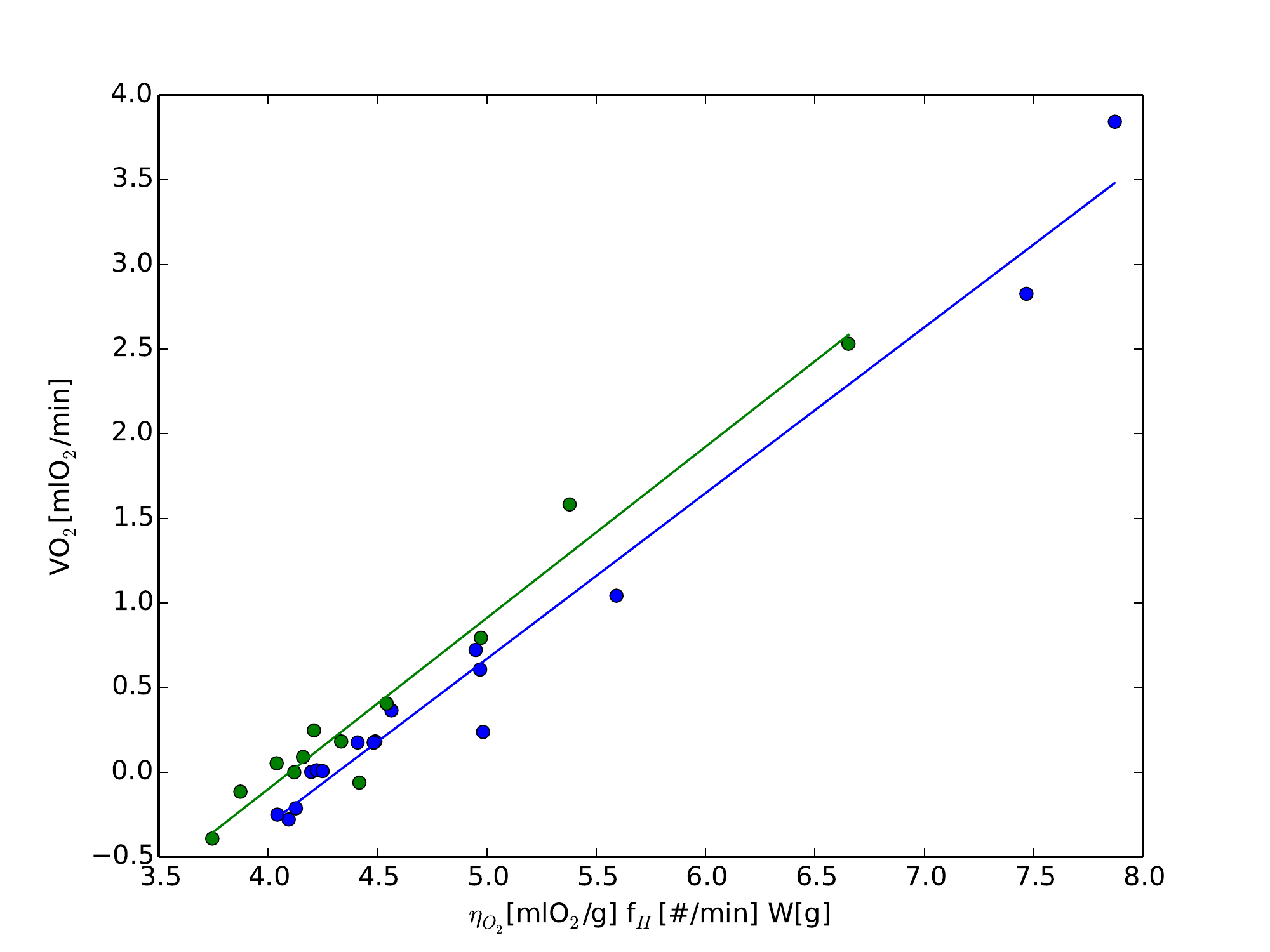}
\caption{Basal metabolic rate $\rm \dot{V}_{O_2} $ as a function of the heart frequencies and mass (with  $\rm \eta_{O_2} = 1 [ml O_2/g]$). The blue points correspond to mammals   and green ones to birds data in logarithmic scale. The curves correspond to the least square fit to the data points, being 0.98 the slope or  mammals and 1.01 the slope for birds.}
\label{F1}
\end{center}
\end{figure}

Blue data points  in Fig \ref{F1} are   the basal metabolic rates  $\rm \dot{V}_{O_2} $  for  mammals and the green ones, $\rm \dot{V}_{O_2} $ for birds taken from Savage et al (2004)  and  Lasiewski \& Dawson (1967) respectively. The respiration rates for both samples  were taken from Calder (1968), converted to heart rates ($\rm f_H$) by multiplying a factor 4.5 for mammals and 9 for birds, values that were taken from the correlations between both rates listed in  Schmidt-Nielsen (1984). The curves correspond to the least square fit to the data points, being 0.98 the slope of the $\rm \dot{V}_{O_2}$  for  mammals and 1.01 the slope of the $\rm \dot{V}_{O_2}$ for birds. These slopes are both close to unity as expected to fulfill dimensional homogeneity. The normalization for birds slightly  differs from mammals, however, this change in  $\rm \dot{V}_{O_2}$ normalization can  be interpreted in our model  
as evidence in variations in the  specific $\rm O_2$ absorption factor $\rm \eta_{O_2}$, since  in Fig \ref{F1} we assumed $\rm \eta_{O_2} = 1 [ml O_2/g]$ for both birds and mammals.  

We  choose the heart rate as characteristic frequency ($\rm  \nu_o = f_H$) instead of the respiration frequency, simply because using  the respiration frequency as $\rm  \nu_o$ increases  the displacement  in normalization  seen between birds and mammals (blue  and green  curves in Fig \ref{F1}),  suggesting   that a formulation with $\rm  \nu_o = f_H$ will  require  less parameters to reach to an unique relation. 
Nevertheless, this argument in favor of $\rm f_H$  do not discard that the respiration frequency could be the controlling physiological parameter.


\begin{figure}[h!]
\begin{center}
\includegraphics[width=11.9cm]{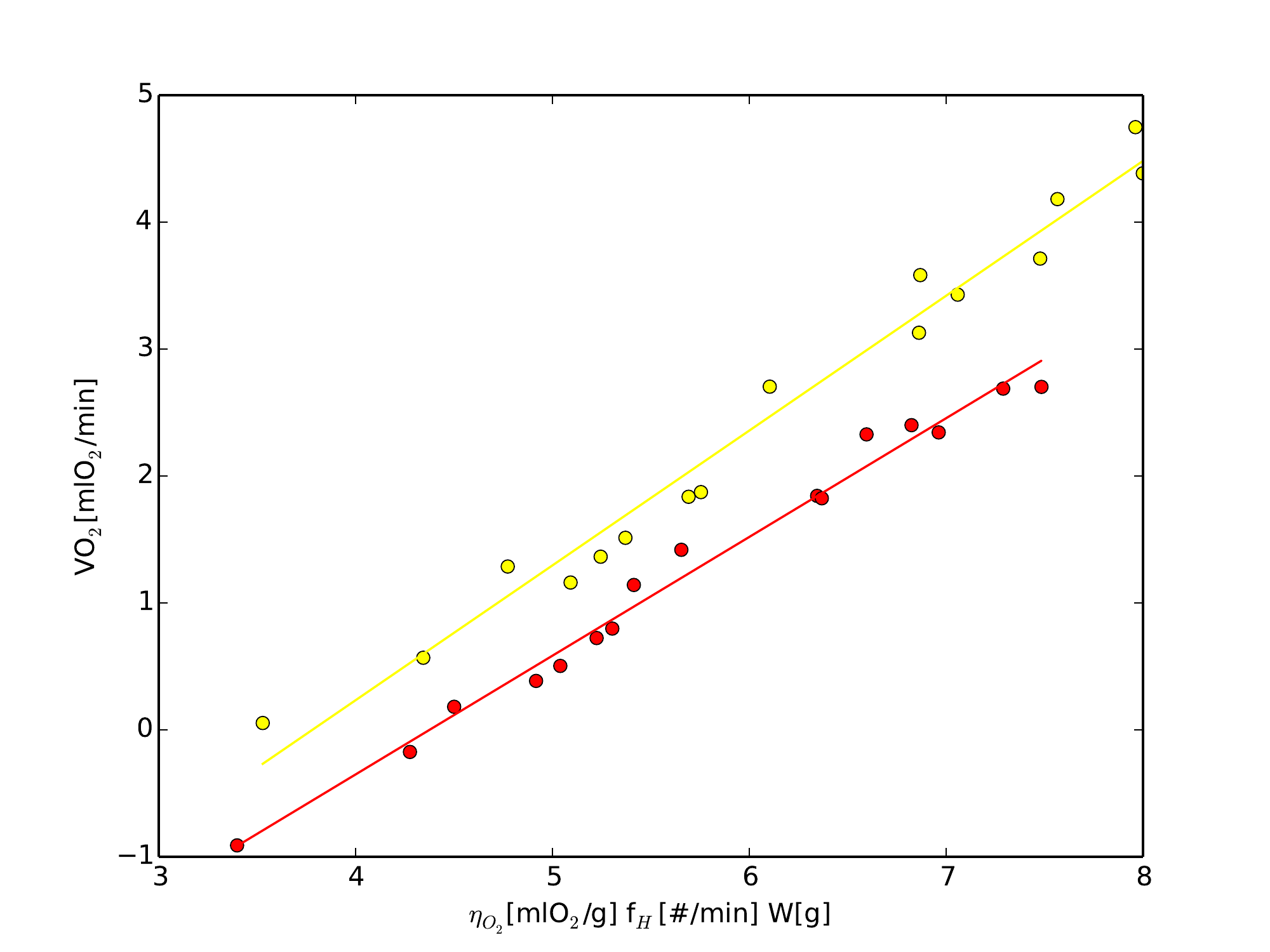}
\caption{Same as Figure \ref{F1}, but  the yellow points  correspond  to maximal metabolic rates $\rm \dot{V}_{O_2}^{max}$ and the red ones are the corresponding  $\rm \dot{V}_{O_2}$ for the same mammals  under basal conditions. The curves correspond to the least square fit to the data points, being 0.94 the slope of the $\rm \dot{V}_{O_2}$  and 1.06 the slope of the $\rm \dot{V}_{O_2}^{max}$. Both axes  are  in logarithmic scale.}
\label{F2}
\end{center}
\end{figure}


Another major  issue in the metabolic rate relation is the  change in slope of the allometric W scaling   when an animal exercises, in the change from $\rm \dot{V}_{O_2}$ to $\rm \dot{V}_{O_2}^{max}$, as mentioned earlier in the text. The red points in Fig. \ref{F2} display the $\rm \dot{V}_{O_2}$ in a sample of resting mammals 
and the yellow ones are the corresponding  $\rm \dot{V}_{O_2}^{max}$ for the same species under maximal exercise.  The sample of mammals was taken from Weibel \& Hoppeler (2004), a group  that has measured  maximal and resting heart frequencies, with the corresponding maximal and basal metabolic  rates taken mainly from Weibel, Bacigalupe et al (2004) and complemented with other references (Hinds et al 1993; Weibel 2000; Roef et al. 2002;  Savage et al 2004; White et al 2006). The curves correspond to the least square fit to the data points, being 0.94 the slope of the $\rm \dot{V}_{O_2}$  and 1.06 the slope of the $\rm \dot{V}_{O_2}^{max}$.  

The larger deviation from slope unity can be explained in terms of the individual slopes of the sample of metabolic rates and frequencies  showed in Fig. \ref{F2}, in which $\rm \dot{V}_{O_2}$ scales as $\rm W^{0.69}$ and the frequency as $\rm f_H \propto W^{-0.26}$,  giving  an $\rm \dot{V}_{O_2}$/$\rm f_H$ proportional to $\rm W^{0.95}$.  Similarly, the $\rm \dot{V}_{O_2}^{max}$    scales as $\rm W^{0.90}$ and  $\rm f_H^{max} \propto W^{-0.15}$ that gives  an $\rm \dot{V}_{O_2}^{max}$/$\rm f_H^{max}$ proportional to $\rm W^{1.05}$ instead of proportional to $\rm W^{1.0}$, as  expected  from  Eq. \ref{new}  for both running and resting conditions, to fulfill  dimensional homogeneity. These variations from slope unity in the running and resting slopes are therefore due to the lower number statistics, because represent only a subsample of animals  that has the three variables measured, in which the metabolic rates allometric exponents  differs  from  the  most accepted values ($\rm  \dot{V}_{O_2} \propto W^{0.75}$ and $\rm  \dot{V}_{O_2}^{max}  \propto W^{0.85}$).
 On the other hand, this support the multiplication/division of averages slopes used in \S 3, since  the observed ones  in  Fig. \ref{F2} are almost the same  slopes expected from both $\rm \dot{V}_{O_2}$/$\rm f_H$ and $\rm \dot{V}_{O_2}^{max}$/$\rm f_H^{max}$. This further support the role of the heart frequency, $\rm f_H$, as an independent physiological variable that controls the metabolic rate.

The shift in normalization seen in  Fig \ref{F2}, can  be interpreted  as a change  in the $\rm O_2$ absorption factor per unit mass, $\rm \eta_{O_2}$ in Eq. \ref{new},  which might  be expected  due to the shift in the animal's internal oxygen  demands between rest and exercise (Darveau et al. 2002; Weibel \& Hoppeler 2005). Under basal conditions, the $\rm {O_2}$ consumption in celular respiration  is mainly determined by the energy demands of basic maintenance processes in the tissues, compared to when an animal exercises that muscle work  places a much larger $\rm {O_2}$ demand for  energy supply (Weibel 2002). This change in the basic processes and organs that controls the energy demands of  animals, should therefore translate into   a shift in their overall  $\rm \eta_{O_2}$ 
 value, which in  Fig \ref{F2}  corresponds   approximately to change in a factor of 5.

In order to quantify the change of $\rm \eta_{O_2}$  from basal to maximal  exercise conditions, we estimate  the   $\rm O_2$ absorption factor per unit mass  for the  species under maximal exercise. Since only the product of $\rm \epsilon \times \eta_{O_2}$  can be extracted from the figures, we will focus on the relative changes by assuming  $\rm \eta_{O_2} = 1 \, [ml O_2/g]$   for resting animals  (blue and green points in Fig \ref{F1} and red ones in Fig \ref{F2}) and then search for the $\rm \eta_{O_2}^{max}$  (yellow points in Fig \ref{F2}) that minimizes the scatter in the relation. An   absorption factor per unit mass of $\rm \eta_{O_2}^{max} = 4.76 \,[ml O_2/g]$ minimizes the scatter of the overall sample, being in such case   0.22 dex with respect to the relation:
\begin{equation}
\rm  log \frac{\dot{V}_{O_2}}{[ml O_2/min]}  =   log \frac{\eta_{O_2}}{[ml O_2/g]}  + log  \frac{f_H}{[\#/min]} + log \frac{W}{[g]}  - 4.313 \, .
\label{fit}
\end{equation}

\begin{figure}[h!]
\begin{center}
\includegraphics[width=11.9cm]{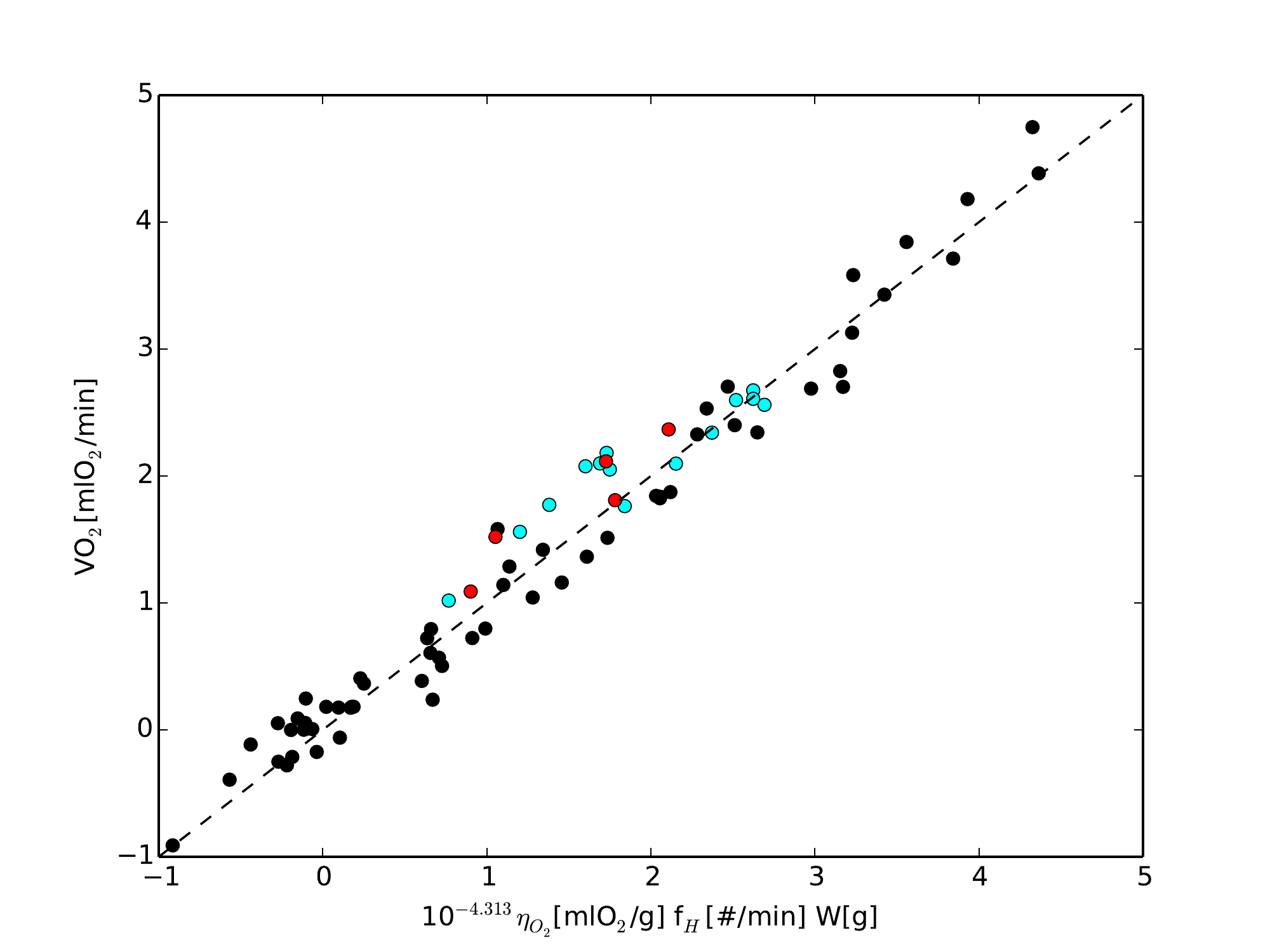}
\caption{Black points are the same data showed  in  Figure \ref{F1} and  \ref{F2}, but we now assume  $\rm \eta_{O_2} = 1 \, [ml O_2/g]$   for resting animals and $\rm \eta_{O_2} = 4.76 \, [ml O_2/g]$ for  those under maximal exercise. The red points correspond to flying birds and the  cyan  ones to penguins under resting and running conditions,  being both subsamples  
within the scatter of the  relation (black points). The black dashed line  denotes the identity between x and y-axes (both in logarithmic scale).}
\label{F3}
\end{center}
\end{figure}


 Fig \ref{F3} displays all the data in Fig  \ref{F1} and   Fig \ref{F2} with black points,  with the just mentioned two different  $\rm O_2$ absorption factor, namely, $\rm \eta_{O_2} = 1 \, [ml O_2/g]$   for resting animals  and $\rm 
 4.76 \, [ml O_2/g]$ for   those under maximal exercise. 
Also, the x-axes is re-normalized using the constant taken from the best fitted model (Eq. {\ref{fit}). This leads to a single relation   consistent with a slope unity,  being an identity relation like  the black dashed line displayed  in Fig \ref{F3}. 

To check  the relation found in Fig \ref{F3}, with a different   $\rm \eta_{O_2}$ for resting and running conditions, we also plotted data for two additional classes of animals. In cyan, we plot data for penguins  under resting and running conditions, taken from Green et al 2005, and for flying birds in red (Berger, Hart \& Roy 1970; Aulie 1971; Norberg 1996). Both subsamples, in which  we again use  $\rm \eta_{O_2} = 1  \,[ml O_2/g]$ for resting subsamples and $\rm \eta_{O_2} = 4.76  \,[ml O_2/g]$ for  exercising  ones, are well described by  the   best fitted model given by Eq. {\ref{fit}, since both the cyan and red points lie within the scatter of the relation (black points). 

For the introduced parameter  $\rm \eta_{O_2}$  is hard to find  direct measurements  in the literature. Nevertheless, one possible estimation for $\rm \eta_{O_2}$ can be taken from the arteriovenous oxygen content difference, $\rm C_{AO_2}$-$\rm C_{VO_2}$, since it plays in the Fick's formula a role equivalent  to $\rm \eta_{O_2}$ in  Eq. \ref{new}. Assuming that the cardiac stroke volume is linearly  proportional to its body mass W (Bishop \& Butler 1995), it is possible to recover from  the Fick's formula an equation of the  form of Eq. \ref{new}, but with a normalization dependent on $\rm C_{AO_2}$-$\rm C_{VO_2}$ instead of $\rm \eta_{O_2}$. 

Factors of two difference in $\rm C_{AO_2}$-$\rm C_{VO_2}$ are in fact reported between resting and exercising/flying birds ({\it Columba Livia}) in Bishop \& Butler (1995), between non-athletic and  athletic mammals (Weibel \& Hoppeler 2005) and even in the case of humans, only  3-months  of moderate physical training increases the maximal oxygen intake  $\rm \dot{V}_{O_2}^{max}$ by 22.5\%, mostly due to a change in  $\rm C_{AO_2}$-$\rm C_{VO_2}$ (Detry et al. 1971). Taking these intra-specific differences into account, an overall  factor of 5    change in $\rm \eta_{O_2}$    for different  classes of animals could be even modest, specially  noting that some  species are intrinsically more aerobically evolved  than others.

\section{Discussion and Outlook}

We showed that although in several  cases the mass scaling in the metabolic rate  and frequency  differs from the  predicted by the `universal 1/4 scaling',  their slopes are consistent with the same  Hausdorff dimension D, if it is interpreted in terms of fractal geometry. This motivate us to propose an unique homogeneous equation for the metabolic rates that includes the  characteristic (heart) frequency  $\rm \nu_o$ as independent controlling variable, for different classes of animals  and for both resting and exercising  conditions, which is in agreement with the empirical data. Our results demonstrate that most of the differences found in the allometric exponents (White et al. 2007) are due to compare incommensurable  quantities, because the variations in the dependence of the metabolic rates on body mass are secondary, coming from variations in the allometric dependence of the heart frequencies on W. Therefore,  $\rm \nu_o$ can be seen as  a new independent physiological variable that controls the metabolic rates, in addition  to the body mass W. 

The  methodology presented in this work could  have 
 impact not only on the field of allometry, but also in life sciences in general. For example, the geometry of body cooling 
that is part of the debate in theories of metabolic allometry from even before Kleiber's result, with our new formulation has no role, thus narrowing the discussion to the internal respiratory ($\rm O_2$) transport system. We consider that our result illustrates how powerful is to apply the $\rm \Pi$ theorem in the formulation of empirical biological laws, something that could also help to validate or discard theories in related  areas such as ecology.


An open question is why the characteristic frequency changes its W scaling among different classes of living organisms and how this depends on the $\rm O_2$ transport problem. The reason for this might still has its origin on  a `fractal like' transport network, however,  the dynamic scaling changes for $\rm \dot{V}_{O_2}^{max}$ should  also  be partially related to  changes in the energy demands (Darveau et al. 2002; Weibel 2002). Since $\rm O_2$ is a crucial ingredient  in the cellular  respiration and energy  storage via  ATP formation,  how this is distributed along  the body and where ends up being consumed, must both play  a role in the integrated  energy consumption of any organism.


A simple separation  of both effects in terms of the complete similarity model presented here, might be to associate the change in the allometric scaling of $\rm \nu_o$ to a change in the transport network itself (which might be fractal  as suggested by West et al 1997 and others, or not, as in Banavar et al 2010), and the shift in normalization ($\rm \eta_{O_2}$) to the change  of the organs that dominates the energy demand (from those associated to  basic maintenance processes to those in muscle work, as in Darveau et al. 2002). 

The evidence  that cells themselves  have constant metabolic rates under `in vitro' conditions (Gauthier et al. 1990), compared to the basal metabolic rates per cell that scales as $\rm W^{-0.25}$ under `in vivo' conditions (for a fixed cell mass; West et al. 2002),  can be interpreted as the allommetric scaling  of $\rm \nu_o$    be only related to the transport system. Moreover, even the case that dynamic changes in W scaling  for $\rm \dot{V}_{O_2}^{max}$ are due to  changes on the  organs that controls the energy demands of animals, this also implies a change  in the transport network itself: from one reaching the organs of basic maintenance processes in the tissues, to one reaching where muscle work places, which in principle should have a different W scaling  for $\rm \nu_o$ .



An interesting  limiting case for   $\rm \nu_o$ is the  metabolic scaling  in plants, since they have (dark) respiration rates (consumption of $\rm O_2$ regardless the presence  of light) linearly proportional to W (Reich et al. 2006). In this context $\rm \nu_o$ corresponds to a  respiration rate and for our model, a metabolic scaling linearly proportional to W  implies a characteristic frequency $\rm \nu_o$ independent of mass.  
One simple interpretation  is since most plants have relatively few `living' cells that carry out cellular respiration outside of their surface, they do not suffer  the transport requirement and therefore such cells are in a condition more similar to the `in vitro' one (Gauthier et al. 1990), having a  characteristic frequency  independent  of the body mass. 

It is important to note that  other theories of metabolic scaling, which are not based on any transport of nutrient supply like the ones discussed so far, are also able to explain the Kleiber's Law. An  example of those is  the Dynamic Energy Budget (DEB) theory (Kooijman 1986, 2000, 2010), that is based on the energetics of reserve dynamics. It can be  shown mathematically,  that this theory    predicts  the same metabolic  rate formulae as in  the West et al (1997) transport model, with only  differences on how the constants depends on the physical parameters of each model (Savage, Deed \& Fontana 2008; Maino et al. 2014).

Under the formulation presented in this work, these different theories  predicts a  different characteristic frequency $\rm \nu_o$ in terms of their model's  physical parameters and in principle,  different predictions of its W scaling for different classes of animals or exercising conditions. This could  be a future test that helps  to discriminate among  theories. It is also an example of how well formulated empirical laws are a more powerful  tool  for  scrutiny   and testing  a  theory, giving additional requirements to be fulfilled. 

Besides the successes  of the relation found, other possible deviations are still documented  in the literature. One particularly relevant is related to the temperature dependence, which is more obvious for Ectothermic systems since their body temperatures strongly varies  with the environmental one. Gillooly et al. (2001) proposed a temperature corrected normalization  for the Kleiber's  metabolic rate relation, based in the  Arrhenius empirical formula for the temperature dependence of chemical reaction rates. 

However, the temperature corrected metabolic relation still has residual variations around of factors 4-5  in the normalization of  the   rates  between endotherms  and ectotherms (Gillooly et al. 2001;  Brown et al 2004). This  is approximately  the same variation in heart rates reported between endotherms (mammals and birds) and ectotherms (fish, anphibians, reptiles)  in Lillywhite et al. (1999), suggesting that Eq \ref{new} with the normalization corrected by an Arrhenius-type exponential dependence on temperature 
 (to take into account that chemical reactions proceed faster at higher T), might be enough to also accurately  account for  the metabolic rates of  ectotherms, using the same relation as for endotherms. 


In terms of dimensional analysis, this can be derived assuming two additional  variables T and $\rm T_a$ that controls the metabolic rate, being $\rm  T_a $ 
 an activation temperature defined as the activation energy $\rm  E_a$ divided by the Boltzmann constant $\rm k_B$ ($\rm  T_a \equiv E_a/k_B$). In such a case, we have now  a  model with n=6 independent 
  variables ($\rm \dot{V}_{O_2} $, $\rm \eta_{O_2}$, W, T,   $\rm T_a$,  $\rm \nu_o$)   that has now only k=4 independent units ($\rm [M_{O_2}]$, [M], [T], $\rm [\Theta]$), therefore, n-k=2 dimensionless parameters can be constructed ($\rm \Pi_1=\dot{V}_{O_2} /W \eta_{O_2} \nu_o $, $\rm \Pi_2=T/T_a$). In this case, the $\rm \Pi$ theorem states that there is an equation f($\rm \Pi_1 $, $\rm \Pi_2$) = 0 and if the function f is regular and differentiable, we can use the implicit function theorem to advocate the existence of a function $\rm \Pi_1=\epsilon (\Pi_2)$. Unfortunately, the functional dependence of $\rm \epsilon$ on the second dimensionless parameter $\rm \Pi_2$ cannot be determined by dimensional analysis,  therefore assuming an exponential form inspired in the empirical Arrhenius formula, namely $\rm \epsilon (\Pi_2) = \epsilon_0  \, {\large e^{\small -1/ \Pi_2}}$, we get

\begin{equation}
\rm \dot{V}_{O_2}  =  \epsilon_0 \, \eta_{O_2} \nu_o  \, {\large e^{\small -E_a/k_BT}} \, W  \, .
\label{finalife}
\end{equation}

This equation can also be used to explain ecological phenomena and make predictions in a similar fashion as Gillooly et al. (2001) uses the temperature corrected  relation on the metabolic theory of ecology (Brown et al 2004). To properly perform  this, all the laws relevant to explain an ecological  phenomena needs to be formulated  in  a form independent of the scaling of units (i.e. dimensionally homogenous equations in their various units of measurements). Once that is achieved, dimensional analysis can also be applied to solve  ecological problems  in a similar  way as engineering does it with physical laws (Bridgman 1937). 

The lack of well formulated laws in life sciences is probably the ultimate origin  of the problems faced by theories  such as  the metabolic theory of ecology, which  produces inaccurate  statements  (universal 1/4 power scaling; White et al 2007) and ecological implications (Duncan et al 2007). Well formulated empirical laws should precede  theory, as happened historically in physics and chemistry. Only after their fundamental laws were formulated thru homogeneous equations in their various units of measurement, and without exceptions,  physics and chemistry  reached the level of predictability characteristic of the exact sciences. 

 
Nevertheless, is still  possible to discuss  implications of Eq. \ref{finalife}  for the few ecological  relations formulated also in dimensionally homogenous equations.  One of such is the population growth, which is an  exponential  controlled by the `Malthusian parameter' or intrinsic rate of natural increase,  $\rm r_m$, which has inverse time units and therefore should be  associated to a frequency. Since its allometic scaling is $\rm r_m \propto W^{-0.25} $ (Fenchel 1974), is natural to associate it in our model with the characteristic (heart) frequency  under basal conditions ($\rm  \nu_o \propto W^{-0.25} $). However, less obvious  is to find a causal connection between two frequencies that represents very different processes and  timescales (internal circulation versus population growth). A possible  link is in the total number of  heartbeats in a lifetime N, which  is approximately constant and equal  to a billion for different mammals  (Cook et al. 2006; Jensen  et al. 2013; Boudoulas et. al. 2015), then if  the  lifetimes   scale inversely  to the  `Malthusian parameter', we have: $\rm  r_m \propto    1/t_{life} \sim \nu_o/N \propto W^{-0.25}$. Animal lifetime is  a more natural timescale  for  controlling  the 
population growth, compared to the inverse of heart frequency. Animal lifetime has also the same allometric scaling of population cycles (Calder 1983). 

A constant  total number N of heartbeats in a lifetime  for mammals, $\rm  t_{life} = N/\nu_o $, can also be used to relate  the normalization $\rm A^+_{ls}$ in the relation for the total  energy consumed in a lifespan (Atanasov 2007), to our normalization ($\rm \epsilon \, \eta_{O_2}$ in Eq \ref{new}). For N equals to a billion and converting 1 ltr $\rm O_{2}$= 20.1kJ (Schmidt-Nielsen 1984), the $\rm A^+_{ls}=7.158 \times 10^5 kJ/kg$  determined  by Atanasov (2007) implies an $\rm \epsilon \, \eta_{O_2} = A^+_{ls}/N =10^{-4.45} \,mlO_2g^{-1}$. This is about the same number that can be determined independently from Fig \ref{F3}, where the normalization of the relation is $\rm
 \epsilon \, \eta_{O_2}  =
 10^{-4.313} \,mlO_2g^{-1}$ (noting  that $\rm \eta_{O_2} = 1 \, [ml O_2/g]$ 
  for resting samples like those in  Atanasov 2007). 
  
Moreover, since in our model the W dependence of the metabolic rate and the heart frequency  $\rm  \nu_o$ are not independent, under  the hypothesis of a constant total number of heartbeats per lifetime,   Eq \ref{new}  predicts that the metabolic rate and the  lifespan are correlated in  such way that the  total  energy consumed in a lifespan  scales linearly with W, regardless  of variations in their exponents. This is indeed observed, for example, in fishes with  metabolic rates (temperature corrected) that scales as  $\rm W^{0.84}$ and the lifespan as $\rm W^{0.16}$, therefore, the  total  energy consumed in a lifespan  scales linearly with W  (Ginzburg and Damuth 2008),  as expected in our model ($\rm =   \epsilon \, \eta_{O_2}  W \, N \propto W$). This result is generally found for poikilothermic animals  (Atanasov 2005a), birds (Atanasov 2005b) and mammals (Atanasov 2007), in spite of variations in metabolic exponents among species. The latter result  also suggest that hypothesis of a constant 
total number of heartbeats in a  lifetime, is valid in other classes of animals besides mammals.
  
We have being able to link  allometric relationships for two different aspect of metabolism, the basal metabolic rate 
with the total energy consumed in a lifespan 
into a single framework, in analogous fashion as Newtonian gravity is able to describe  free-fall acceleration on different planets 
in \S 2.  Only  with the assumption  of a constant  total number N of heartbeats in a lifetime, we found  consistent  dimensional constants values $(\rm \epsilon \, \eta_{O_2} \sim A^+_{ls}/N)$ and  consistent mass scaling variations due to  secondary variables
, showing again the advantages of using empirical relations that are mathematically well defined (both relationships fulfill homogeneity). In this case the exceptions (metabolic rates, heart frequencies and lifespans departing from the  universal `1/4 power law' mass scaling), which are characteristic of the empirical relations in the so-called 
 `inexact sciences'
, was due to an inaccurate  mathematical formulation of the relationship  and not due to intrinsic higher complexity of the problem. Therefore, it seems promising that more  relationships in Biology/Ecology could 
be reformulated in a dimensionally homogenous form. 


Finally, for further  fine-tuning  of the metabolic rate relation, a coherent dataset of  measurements is still required. Ideally, this needs to be  for all the variables   measured on the same animals and  the original  formulation  of the metabolic law should stay strictly with measured quantities, for example with metabolic rates in ml $\rm O_2$ per min instead of  energy-related units (watts or ergs/sec), to avoid  assumptions in the  conversion factors that  produces  undesirable extra scatter in the relation. Also, it will be  interesting to test its predictions, for example to look for the temporal validity of Eq. \ref{finalife} as an animal starts exercising  and  increases its  $\rm O_2$ consumption. In ecology, an interesting  test is to see if  (Malthusian) population ecology parameters scale as $\rm \nu_o $,  instead of the universal 1/4 expected in other approaches in metabolic ecology (Ginzburg and Damuth 2008). 
A good candidate  to see changes in allometic scaling of $\rm r_m$, are big  outliers  from `1/4 scaling' in the metabolic relation under basal conditions, such as spiders  (Anderson 1970, 1974) or other organisms (White et al 2007).
 


I thank  Leonardo Bacigalupe for his help with  the heart rates presented in Weibel \& Hoppeler (2004) and   the corresponding metabolic rates in Weibel, Bacigalupe et al (2004). I also acknowledge partial support from the Center of Excellence in Astrophysics and Associated Technologies (PFB 06).


\begin{thebibliography}{}

\bibitem[Binney \& Tremaine (1994)]{jre45} Anderson, J.F. (1970). Metabolic rates of spiders, Comp. Biochem. Physiol., 33, 51-72
\bibitem[Kennicutt (1998)]{ken01} Anderson, J.F. (1974). Responses to Starvation in the Spiders Lycosa Lenta Hentz and Filistata Hibernalis (Hentz), Ecology, 55, 576-585
\bibitem[Kennicutt (1998)]{ken01} Atanasov, A.T., 2005a. The linear allometric relationship between total metabolic energy per life span and body mass of poikilothermic animals. Biosystems 82, 137-142.
\bibitem[Kennicutt (1998)]{ken01} Atanasov, A.T., 2005b. Linear relationship between the total metabolic energy per life span and the body mass of Aves. Bulgarian Med. XIII, 30-32.

\bibitem[Kennicutt (1998)]{ken02} Atanasov,  A. T.  (2007). The Linear Allometric Relationship Between Total Metabolic Energy per Life Span and Body Mass of Mammals, BioSystems, 90, 224-233.
\bibitem[Binney \& Tremaine (1987)]{ba88} Aulie, A. (1971). Co-ordination between the activity of the heart and the flight muscles during flight in small birds, Comp. Biochem. Physiol., 38, 91-97
\bibitem[Binney \& Tremaine (1987)]{az98} Bainbridge, R., (1958). The speed of swimming as related to size and to the frequency and amplitude of the tail beat, J. Exp. Biol., 35, 109-133.
\bibitem[Downes \& Solomon (1998)]{ds98} Barenblatt, G. I. \& Monin, A.S. (1983). Similarity principles for the biology of pelagic animals, Proceedings of the National Academy of Sciences USA 80, 3540-3542.
\bibitem[Downes \& Solomon (1998)]{de98} Barenblatt, G. I. (2003). {\it Scaling}, Cambridge University Press 
\bibitem[Kennicutt (1998)]{ken89} Berger, M., Hart, J.S., \& Roy, O.Z. (1970). Respiration, Oxygen Consumption and Heart Rate in Some Birds during Rest and Flight, Z. vergl. Physiologic, 66, 201-214
\bibitem[Binney \& Tremaine (1994)]{gi98} Bishop, C.M. (1999). The maximum oxygen consumption and aerobic scope of birds and mammals: getting to the heart of the matter, Proceedings of the Royal Society of London B 266, 2275-2281
\bibitem[Binney \& Tremaine (1987)]{az92} Bolster, D. et al. (2011). Dynamic similarity: the dimensionless science, Physics Today, 64, 42-47
\bibitem[Binney \& Tremaine (1987)]{az92} Boudoulas, K.D. et. al (2015). Heart Rate, Life Expectancy and the Cardiovascular System: Therapeutic Considerations, Cardiology, 132, 199-212
\bibitem[Binney \& Tremaine (1987)]{bk88}  Bridgman, P. W. (1922). {\it Dimensional Analysis}, Yale University Press
\bibitem[Binney \& Tremaine (1994)]{la81}  Brody, S. (1945). {\it Bioenergetics and Growth}, Reinhold, New York.
\bibitem[Binney \& Tremaine (1987)]{bk88}    Brown, J. et al. (2004). Toward a Metabolic Theory of Ecology, Ecology, 85, 1771-1789
\bibitem[Binney \& Tremaine (1987)]{bt87} Buckingham, E. (1914). On Physically Similar Systems: Illustrations of the Use of Dimensional Analysis, Physical Review, 4, 345-376
\bibitem[Binney \& Tremaine (1994)]{el044} Calder, W.A. (1968) III Respiratory and heart rates of birds at rest. Condor 70, 358-365
\bibitem[Binney \& Tremaine (1994)]{dr99} Calder, W.A. (1983).  An Allometric Approach to Population Cycles of Mammals, J. theor. Biol., 100, 275-282 
\bibitem[Binney \& Tremaine (1994)]{el02} Carrel, J.E., Heathcote, R.D. (1976). Heart Rate in Spiders: Influence of Body Size and Foraging Energetics, Science, 193, 148-150
\bibitem[Binney \& Tremaine (1987)]{az05}  Cook, S. et al. (2006). High heart rate: a cardiovascular risk factor? European Heart Journal, 27, 2387-2393
\bibitem[Kennicutt (1998)]{ken98} Darveau, C.A., Suarez, R.K., Andrews, R.D., Hochachka, P.W., (2002). Allometric cascade as a unifying principle of body mass effects on metabolism, Nature, 417, 166-170
\bibitem[Downes \& Solomon (1998)]{el08} Detry, J.M. et al. (1971). Increased Arteriovenous Oxygen Difference After Physical Training in Coronary Heart Disease, Circulation, 44, 109-118
\bibitem[Downes \& Solomon (1998)]{el08} Dodds, P.S., Rothman, D.H. \& Weitz, J.S., (2001). Re-examination of the 3/4-law of Metabolism, Journal of Theoretical Biology , 209, 9-27 
\bibitem[Downes \& Solomon (1998)]{el08} Dlugosz, E.M. et al (2013). Phylogenetic analysis of mammalian maximal oxygen consumption during exercise, The Journal of Experimental Biology, 216, 4712-4721
\bibitem[Binney \& Tremaine (1994)]{ga05} Duncan, R.P. et al. (2007). Testing  The Metabolic  Theory  of Ecology: Allometric Scaling  Exponents in Mammals, Ecology, 88, 324-333 
\bibitem[Binney \& Tremaine (1994)]{ke05} Escala, A. (2015). On the Functional Form of the Universal Star-formation Law, The Astrophysical Journal,  804, 54-62
\bibitem[Binney \& Tremaine (1994)]{pe4} Fenchel, T. (1974). Intrinsic Rate of Natural Increase: The Relationship with Body Size, Oecologia,14, 317-326
\bibitem[Binney \& Tremaine (1994)]{gl65}  Fourier, Joseph (1822). {\it Th\'eorie analytique de la chaleur}, Firmin Didot, Paris
\bibitem[Kennicutt (1998)]{ko05} Gauthier, T., Denis-Pouxviel, C. \& Murat, J. C. (1990). Int. J. Biochem. 22,
411-417
\bibitem[Binney \& Tremaine (1994)]{he08} Gillooly, J.F., Brown, J.H., West, G.B., Savage, V.M. \& Charnov, E.L. (2001). Effects of size and temperature on metabolic rate, Science, 293, 2248-2251.
\bibitem[Binney \& Tremaine (1994)]{he08} Ginzburg, L., Damuth, J. (2008) The Space-Lifetime Hypothesis: Viewing Organisms in Four Dimensions, Literally, American Naturalist 171, 125-131
\bibitem[Binney \& Tremaine (1987)]{az98} Green, J.A. et al (2005). Allometric estimation of metabolic rate from heart rate in penguins, Comparative Biochemistry and Physiology Part A, 142, 478-484
\bibitem[Binney \& Tremaine (1994)]{he03} Hinds, D.S. et al. (1993). Maximum  Metabolism  \& The Aerobic factorial scope of endotherms, J. exp. Biol., 182, 41-56
\bibitem[Binney \& Tremaine (1994)]{rpe4}  Hulbert, A. J. (2014). A Sceptics View: `Kleiber's Law' or the `3/4 Rule' is neither a Law nor a Rule but Rather an Empirical Approximation, Systems, 2, 186-202
\bibitem[Binney \& Tremaine (1994)]{ke87} Jensen MT, et al. (2013). Elevated resting heart rate, physical fitness and all-cause mortality: a 16-year follow-up in the Copenhagen Male Study, Heart, 99, 882-887
\bibitem[Binney \& Tremaine (1994)]{ko01} Kleiber, M. (1932). Body size and metabolism, Hilgardia, 6, 315-351
\bibitem[Binney \& Tremaine (1994)]{ko09} Kooijman, S.A.L.M. (1986). Energy budgets can explain body size relations. Journal of Theoretical Biology, 121, 269-282.
\bibitem[Binney \& Tremaine (1994)]{ko00}Kooijman, S.A.L.M. (2000). Dynamic Energy and Mass Budgets in Biological Systems. Cambridge University Press, Cambridge.
\bibitem[Binney \& Tremaine (1994)]{ko10}Kooijman, S.A.L.M. (2010). Dynamic Energy Budget Theory for Metabolic
Organisation. Cambridge University Press, Cambridge.
\bibitem[Binney \& Tremaine (1994)]{jo05} Lasiewski, R.C. \& Dawson, W.R. (1967). A re-examination of the relation between standard metabolic rate and body weight in birds, Condor 69, 13-23.
\bibitem[Binney \& Tremaine (1994)]{la81}  Lillywhite, H.B. et al. (1999). Resting and maximal heart rates in ectothermic vertebrates, Comparative Biochemistry and Physiology Part A, 124, 369-382
\bibitem[Binney \& Tremaine (1994)]{ma14}  Maino, J.L.  et al. (2014). Reconciling theories for metabolic scaling, Journal of Animal Ecology, 83, 20-29.
\bibitem[Binney \& Tremaine (1994)]{re4} Norberg, U. M. (1996). {\it The energetics of flight, pages 199-249 in Avian energetics and nutritional ecology}, Chapman and Hall, New York, USA
\bibitem[Binney \& Tremaine (1994)]{roe4} Price, C.A., Ogle, K., White, E. P.  \&  Weitz, J. S. (2009). Evaluating scaling models in biology using hierarchical Bayesian approaches, Ecology Letters,  12, 641-651
\bibitem[Binney \& Tremaine (1994)]{roe4}  Rayleigh (1915). The Principle of Similitude, Nature, 95, 66-8
\bibitem[Binney \& Tremaine (1994)]{pe4} Reich, P. B. et al. (2006). Universal scaling of respiratory metabolism, size and nitrogen in plants, Nature, 439, 457-461
\bibitem[Binney \& Tremaine (1994)]{sa02} Roef, M.J. et al. (2002). Resting oxygen consumption and in vivo
ADP are increased in myopathy due to complex I deficiency, Neurology, 58, 1088-1093
\bibitem[Binney \& Tremaine (1994)]{sa59} Savage, V. M. et al. (2004). The predominance of quarter-power scaling in biology, Funct. Ecol., 18, 257-282
\bibitem[Binney \& Tremaine (1994)]{so59} Savage, V.M., Deeds, E.J. \& Fontana, W. (2008). Sizing up allometric scaling theory. PLoS Computational Biology, 4, e1000171.
\bibitem[Binney \& Tremaine (1987)]{bt87} Schmidt-Nielsen, K. (1984). {\it Scaling: why is animal size so important?},      Cambridge University Press.
\bibitem[Binney \& Tremaine (1994)]{wn07} Stahl, W.R. (1967). Scaling of respiratory variables in mammals, Journal of Applied Physiology 22, 453-460
\bibitem[Binney \& Tremaine (1994)]{va94} Taylor, G. I. (1950). The Formation of a Blast Wave by a Very Intense Explosion. II. The Atomic Explosion of 1945, Proc. R. Soc. Lond. A, 201, 175-186
\bibitem[Binney \& Tremaine (1994)]{wn09} Taylor, C.R., Maloiy, G.M.O., Weibel, E.R., Lungman, V.A., Kamau, J.M.Z., Seeherman, H.J. \& Heglund, N.C. (1981) Design of the mammalian respiratory system. 3. Scaling maximum aerobic capacity to body-mass: wild and domestic animals, Respiratory Physiology 44, 25-37
\bibitem[Binney \& Tremaine (1994)]{you01} Turcotte, D.L., Pelletier, D.L., Newman, W.I. (1998). Networks with Side Branching in Biology, Journal of Theoretical Biology, 139, 577-592
\bibitem[Binney \& Tremaine (1994)]{to64} Utreras, J., Becerra, F., Escala, A. (2016). Unveiling the Role of Galactic Rotation on Star Formation, The Astrophysical Journal, 833, 13-31	
\bibitem[Binney \& Tremaine (1994)]{wn01} von K\'arm\'an, Th. (1957). {\it Aerodynamics}, Cornell University Press, Ithaca.
\bibitem[e.R. _(Binney \& Tremaine (1994)]{np97} Weibel, E.R. (2000). {\it Symmorphosis: on Form and Function in Shaping Life.} Harvard University Press, Cambridge.
\bibitem[Binney \& Tremaine (1994)]{la82} Weibel, E.R. (2002). The pitfalls of power laws, Nature, 417, 131-132
\bibitem[e.R. _(Binney \& Tremaine (1994)]{np97} Weibel, E.R., Bacigalupe, L.D., Schmidt, B. \& Hoppeler, H. (2004). Allometric scaling of maximal metabolic rate in mammals: muscle aerobic capacity as a determinant factor. Respiration Physiology and Neurobiology, 140, 115-132
\bibitem[e.R. _(Binney \& Tremaine (1994)]{np97} Weibel, E.R. \& Hoppeler, H. (2004). Modeling Design and Functional Integration in the Oxygen and Fuel Pathways to Working Muscle, Cardiovascular Engineering, 4, 5-18
\bibitem[e.R. _(Binney \& Tremaine (1994)]{np97} Weibel, E.R. \& Hoppeler, H. (2005). Exercise-induced maximal metabolic rate scales with muscle aerobic capacity, Journal of Experimental Biology, 208,1635-1644
\bibitem[Binney \& Tremaine (1994)]{ll03} West, G.B., Brown, J.H. \& Enquist, B.J. (1997). A general model for the origin of allometric scaling laws in biology. Science 276, 122-126
\bibitem[Binney \& Tremaine (1994)]{ll03} West, G.B., Brown, J.H. \& Enquist, B.J. (1999). The Fourth Dimension of Life: Fractal Geometry and Allometric Scaling of Organisms. Science 284, 1677-1679
\bibitem[Binney \& Tremaine (1994)]{kk01} West, G.B., Woodruff, W.H. \& Brown, J.H. (2002). Allometric scaling of metabolic rate from molecules and mitochondria to cells and mammals. Proceedings of the National Academy of Sciences USA, 99, 2473-2478
\bibitem[Binney \& Tremaine (1994)]{sa98} West, G., Brown, J. (2004). Life's Universal Scaling Laws, Physics Today, 57, 36-42
\bibitem[Binney \& Tremaine (1994)]{van01} White, Craig R. et al. (2007). Allometric Exponents Do Not Support a Universal Metabolic Allometry, Ecology, 88, 315-323




\end{thebibliography}
\end{document}